\documentclass[preprint,prb,aps]{revtex4}
\begin{document}

\title{RELATION TIME-THERMODYNAMICS. A PATH INTEGRAL APPROACH.} 
\author{J.P. Badiali} 
\address{LECA, ENSCP-Universit\'e Pierre et Marie Curie
4 Place Jussieu, 75230 Paris Cedex 05, France}

\begin{abstract}
Starting from an algebraic approach of quantum physics it has been shown via 
the Tomita-Takesaki theorem and the KMS condition that the canonical density 
matrix contains the dynamics of the system provided we use a rescaling of 
time. In this paper we show that the path integral form of the partition 
function reveals a dynamics which is complementary of the one 
given by the Tomita-Takesaki theorem. To do that we work in the spirit 
of a Feynman'conjecture. We define the entropy as a measure of the 
disorder in space time. By using an equilibrium condition we 
introduce a natural time scale that it is precisely the one appearing in the 
Tomita-Takesaki theorem. For this time scale depending on the 
temperature but not on the system properties our definition of entropy is 
identical to the thermodynamic one. The underlying dynamics associated 
with the partition function allows us to derive a $\bf{H}$-$theorem$. It is 
obtained in the thermodynamic limit and provided we are in a regime in which 
the thermal fluctuations are larger than the quantum ones.

%\\

\end{abstract}
\maketitle

%%%%%%%%%%%%%%%%%%%%%%%%%%%%%%%%%%%%%%%%%%
\vspace{0.5cm}

\section{Introduction}
When a new field of investigation is developing we frequently observe that 
different equivalent approaches are proposed. These approaches focusing 
on different starting points may reveal different and non expected aspects 
of the theory. Moreover, after sometimes new extensions or new 
generalizations of the initial description may be proposed. This was the 
case for classical mechanics (\cite{arnold}) where after the Newton laws 
the Lagrangian and Hamiltonian formalisms have been introduced while, 
more recently, a symplectic approach has been developed. The last one offers 
a scheme in which it is possible to recast the general relativity 
(\cite{rove1}). A similar evolution is true in the case of quantum 
mechanics. Almost at the same time Heisenberg elaborated his matrix 
mechanics and Schroedinger proposed his famous equation, while later Feynman 
introduced the path integral formalism. If all these approaches are 
equivalent at the level of the Schroedinger equation, the path integral 
formalism has been extended far beyond in order, for example, to describe 
the partition function near a black hole (\cite{gibbons}).\\ 
The matrix mechanics has giving rise to the so called algebraic approach of 
quantum mechanics (\cite{emch}) in which the main point is the existence of 
a C* algebra of non-commutative operators. Then due to the GNS construction 
(\cite{emch}) it is possible to give a concrete definition of these 
operators by a representation in a Hilbert space constructed from a given 
state that is called cyclic and noted $\Omega$. An important result of this 
approach is the Tomita-Takesaki theorem (\cite{tomita})(for a recent review 
on this theorem see (\cite{summers})) that establishes the existence of a 
one parameter group of automorphisms of the algebra \begin{equation} 
\alpha_{t}A = \Delta^{-it} A \Delta^{it} 
\label{auto} 
\end{equation} 
that is true for any operator $A$ of the algebra and leaves the algebra  
globally invariant. In (\ref{auto}) $\Delta$ is a self adjoint positive 
operator and $\alpha_{t}$ is the so called modular group on the 
algebra. Another consequence of the Tomita theorem is the existence of a 
relation \begin{equation} 
(\Omega, B(\alpha_{t+i}A)\Omega) = (\Omega, (\alpha_{t}A) B\Omega) 
\label{tomita} 
\end{equation}
in which (,) means the inner product. It has been demonstrated that such 
relation is independent of the particular state but it is an intrinsic 
property of the von Neumann algebra (\cite{connes1}). At this 
level it is impossible to claim that the parameter $t$ in (\ref{tomita}) is 
a time. However we may compare (\ref{tomita}) with the KMS condition 
(\cite{kms}) established by Kubo (\cite{kubo}) and Martin and Schwinger 
(\cite{martin}). Haag and his coworkers (\cite{HHW}) postulated that the KMS 
condition is the correct definition of thermal equilibrium for infinite 
dimensional quantum systems. The KMS condition shows that the correlation 
function between two variables A and B noted $<(\gamma_{t}A)B>$ is analytic 
in the strip $0<Im(t)<\beta\hbar$ where $\beta$ is the usual reverse of 
temperature ($\beta = \frac{1}{k_{B}T}$) and we have \begin{equation}
<(\gamma_{t}A)B> = <B (\gamma_{t+i\beta \hbar}A)> 
\label{kms}
\end{equation}
where $\gamma_{t}$ introduces the time translation group defined according to
\begin{equation}
\gamma_{t} A = \exp{\frac{it H}{\hbar}} A \exp{-\frac{it H}{\hbar}}
\label{evolution}
\end{equation} 
in which $H$ is the hamiltonian operator. Finally we find that the result of the Tomita-Takesaki theorem is equivalent 
to the time evolution of the bounded operators generated by the Hamilton 
provided the time is measured in units $\beta \hbar$ as already noted in 
(\cite{connes}). These works show that the canonical density matrix
\begin{equation}
\rho = \frac{exp-\beta H}{N}
\label{gibbs}
\end{equation}
where $N$ is a normalization constant
contains the dynamics of the system. Connes and Rovelli (\cite{connes}) 
suggested an extension of the previous results to general covariant theories 
via the thermal time hypothesis in which the physical time is given 
by the modular group (\ref{auto}). Hereafter our main task will be to 
explain the origin of the rescaling $ t \to \frac{t}{\beta \hbar}$.\\ 
This paper is organized as follows. In section $2$ we develop a path 
integral approach in the spirit of a Feynman'conjecture. We show that the 
usual partition function involved a given dynamics restricted to time 
interval $t \le \beta \hbar$. Then, from simple physical arguments, we 
introduce an entropy characterizing the disorder in space time. In section 
$3$ we show that a thermal equilibrium condition leads to introduce a time 
interval $\tau = \beta \hbar$ that is independent of the system but 
determined by the thermodynamic state via the temperature. This time interval 
appears as a natural time scale in equilibrium thermodynamics, indeed it is 
precisely the time scale related to the Tomita-Takesaki theorem. Moreover 
our definition of equilibrium is very similar to the one introduced by 
Rovelli (\cite{rove1}). Traditionally time and thermodynamics are related via 
the second law of thermodynamics, which asserts the existence of a state 
function that is a non-decreasing function of time for any closed system 
(\cite{landms}). In statistical mechanics this leads to the problem of the 
arrow of time and the so-called $\bf{H}$-$theorem$ (for a review see for instance 
(\cite{zeh})). In section $4$ we show that the dynamics associated with the 
partition function allows the derivation of a $\bf{H}$-$theorem$, in order to do 
that we define a $\bf{H}$ function showing the interplay between open and close 
paths. In section $5$, we give some conclusions.

\section{Statistical mechanics and path integral formalism}
Starting from (\ref{gibbs}) Feynman (\cite{feynman}) derived a path integral 
formalism for the partition function $Z$. In order to do that it was needed 
i) to start from the wave function, the existence of stationary states and 
eigenvalues, ii) to develop some arguments justifying the use of the 
canonical form of the density matrix and iii) to introduce some mathematical 
tricks. Then the partition function can be written as \begin{equation} 
Z =\int dx \int Dx(t) \exp{-\frac{1}{\hbar} A_{0}[x(t); 0,\tau ]}
\label{Zfunction}
\end{equation}
in which $Dx(t)$ is the measure for the functional integral and 
$A_{0}[x(t);0,\tau]$ represents the quantity 
\begin{equation}
A[x(t);0, \tau] = \int\limits_{0}^{\tau} [\frac{1}{2} m 
[\frac{dx(t)}{dt}]^{2} + V(x(t))] dt] 
\label{action}
\end{equation} 
calculated on loops $i.e.$ on trajectories on which we have $x(t=0)$ = 
$x(t=\tau)$ = $x$; for each loop the dummy variable, $t$, varies from $0$ to 
$\tau$, $t$ and $\tau$ have the dimension of a time. In (\ref{action}) it is 
assumed that the external potential $V(x(t))$ does not depend explicitly on 
$t$. The integral (\ref{Zfunction}) is calculated on the all values of $x$ 
taken in the volume of the system under consideration. The expression of $Z$ 
given in (\ref{Zfunction}) is just a mathematical transformation of the 
standard expression of $Z=Tr(\exp[-\beta H])$ provided we use $\tau=\beta 
\hbar$. Of course in such a formal derivation we cannot claim that $t$ is a 
time having a physical meaning $i.e.$ that $A_{0}[x(u);0, \tau]$ is an 
action associated to loops in ordinary space time.\\ After deriving 
(\ref{Zfunction}) Feynman wrote in (\cite{feynman}) a paragraph entitled 
''Remarks on methods of derivation''on which he suggested that it should be 
possible to derive the expression of $Z$ directly from the path integral 
description for the time-dependent motion how he did for the quantum 
mechanical amplitude in quantum mechanics. Why to search such a short cut 
that should avoid the wave function and the energy levels ? The Feynman's 
answer is the following ''in doing that a deeper understanding of physical 
processes might result or possibly more powerful methods of statistical 
mechanics might be evolved''. Hereafter, to be short we will mention these 
comments as the Feynman's conjecture $(FC)$. From the $FC$ it is suggested 
that $Z=Tr(\exp[-\beta H])$ contains the dynamics of the system provided we 
use its path integral version. This is very similar to what has been exposed 
above where it has been shown that $\exp[-\beta H]$ may reveal the dynamics 
of the system if we use an algebraic approach of quantum mechanics. \\ 
Hereafter, we will try to explicit the dynamics involved in the path 
integral version of $Z$ and to show that $\tau = \beta\hbar$ is the natural 
unit of time in equilibrium thermodynamics. In order to do that the first 
step is to check that all the results obtained via the path integral 
formalism are meaningful if we consider that $t$ is a physical time in the 
action $A[x(t);0, \tau]$. In (\cite{jpb1}) it has been shown that loops are 
fractal trajectories with the same Hausdorff dimension as the one calculated 
from Schroedinger equation. We may also shown, for a free particle, that the 
trajectories are mainly restricted to a sphere for which the radius 
corresponds to the thermal de Broglie wavelength, an expected result. 
Moreover, a first meaning of the time $\tau$ can be given by an analysis in 
terms of time/energy uncertainty relation, $\tau$ appears as the time 
interval we need to wait in order to reach a regime in which the quantum 
fluctuations become smaller than those of thermal origin (\cite{jpb1}), 
(\cite{jpb2}). However these results are not sufficient to give a strong 
basis to the $FC$. In order to go beyond we must show that i) the dynamics 
implicitly involved in (\ref{Zfunction}) is meaningful and ii) the form of 
the functional integral can be written down immediately from simple physical 
arguments.\\

\subsection{Dynamics implicitly involved in $Z$}
From $Z$ we may define the quantity \begin{equation} 
q(t_{0},x_{0};t,x) = \smallint 
Dx(t) \exp -\frac{1}{\hbar} A[x(t); t_{0},t] 
\label{q0}
\end{equation} 
$q(t_{0},x_{0};t,x)$ appears as a weighted sum of all the 
paths $x(t)$ connecting the points $x_{0}$ to $x$ during an interval 
$(t- t_{0})$.
From $q(t_{0},x_{0};t,x)$ and a function $\phi_{0}(x)$ defined 
for $(t = t_{0})$ we may form a real-valued function $\phi(t,x)$ according 
to 
\begin{equation} 
\phi(t,x) = \smallint \phi_{0}(y) q(t_{0},y;t,x) dy 
\label{fi1} 
\end{equation} 
By using the Feynman-Kac formula, we can see that $\phi(t,x)$ is the 
solution of the equation
\begin{equation} 
-{\partial  \phi(t,x)}/{ \partial t} + {\frac{\hbar}{2m} \Delta_{x} 
\phi(t,x)} -\frac{1}{\hbar}{u(t, x) \phi(t,x)} = 0 
\label{dif} 
\end{equation}
that verifies the initial condition $\phi(0,x)= \phi_{0}(y)$ $i.e.$ 
$q(t_{0},x_{0};t,x)$ is the fundamental solution of (\ref{dif}) 
in which $\Delta _{x}$ means the laplacian operator taken at the point $x$.
Note that (\ref{dif}) is not a Chapman-Kolmogorov type equation and 
consequently $q(t_{0},x_{0};t,x)$ is not a density of probability, this 
quantity can not be normalized in general (\cite{naga1}). However 
$q(t_{0},x_{0};t,x)$ verifies the law of composition (\cite{naga1}) 
\begin{equation} q(t_{1},x_{1};t_{2},x_{2}) = \smallint dx_{3} 
q(t_{1},x_{1};t_{3},x_{3}) q(t_{3},x_{3};t_{2}, x_{2}) \label{chapkol} 
\end{equation} provided $t_{1} < t_{3} < t_{2}$ and therefore it can be used 
to describe the transitions in space-time. The evolution in space-time is 
then represented by a semi-group, $\phi(t,x)$ is defined for $t \ge 
t_{0}$ and the equation (\ref{dif}) is time-irreversible.  \\ 
The equation (\ref{dif}) has the classical form of a diffusion equation 
in an external potential, $\frac{\hbar}{2m}$ plays the role of a diffusion 
coefficient. To summarize, if we accept that the parameter $t$ that 
appears in (\ref{Zfunction}) is a physical time then the underlying dynamics 
is given by (\ref{dif}). The next question is: can we accept this dynamics 
as realistic ? In (\cite{jpb3}) we have shown how to justify 
(\ref{dif}) starting from a primarily discrete space-time and to recast 
(\ref{dif}) in general trends in modern physics. The next step is to show 
that we can write thermodynamic properties directly in terms of path 
integral. \\

\subsection{Path integral expression of the entropy}
From standard thermodynamics we know that $Z$ is related to the free energy 
$F$ by $F=-k_{B}T \ln{Z}$; we have also $F=U-TS$ in which $S$ is the 
entropy and $U$ the internal energy. We may consider that $U$ is an external 
parameter fixed by the mode of preparation of the system. From the previous 
thermodynamic relations and (\ref{Zfunction}) we introduce a path integral 
expression of the entropy given by 
\begin{equation} 
S = k_{B} \ln \smallint dx \smallint 
Dx(t) \exp{ - \frac{1}{\hbar} [A_{o}[x(t);0, \tau] - \tau U] }. 
\label{gam} 
\end{equation}
in which we consider temporarily $\tau$ as a free parameter and the quantity 
$\tau U$ as the thermal action. We see that $S$ is 
determined by the difference $A_{o}[x(t);0, \tau] - \tau U$. If it should 
exist only one possible trajectory for which the euclidean action should 
exactly compensate the thermal one should say that there is no disorder in 
the system and from (\ref{gam}) we can see that $S =0$ as expected. For real 
systems it may exist some fluctuations of $A_{o}[x(t);0, \tau]$ and we may 
have a lot of trajectories for which the order of magnitude of 
$A_{o}[x(t);0, \tau] - \tau U$ is approximately $\hbar$, all these 
trajectories contribute to $S$. Larger is this number of trajectories 
smaller is the order in space time and larger is the entropy. From 
(\ref{gam}) we see that the calculation of $S$ requires to start from a 
point $x = x(t=0)$ in space and to explore during a time interval $\tau$ all 
the loops around this point and, finally, to perform the same procedure 
for each value of $x$ in the volume of the sample. It is clear 
that $S$ characterizes the order in space time. Thus from the 
dynamics it is possible to write immediately $S$ from physical arguments as 
suggested by the $FC$. Because there is no explicit Gibbs ensembles in our 
definition of entropy there is no direct relation between (\ref{gam}) and 
canonical or micro-canonical representations of statistical mechanics. In 
fact (\ref{gam}) is an entropy representation of statistical mechanics as 
defined by Callen (\cite{cal}). The question is now to see if such a 
definition of $S$ leads to the thermodynamic result. In order to do that we 
have to find a manner to determine the free parameter $\tau$. \\

\section{The natural unit of time}
The entropy defined above depends on two external parameters 
$\tau$ and $U$ that are not independent. From 
the definition (\ref{gam}) we may calculate the 
derivative $\frac {dS}{dU}$ that is the reverse of the temperature $T$ in 
the usual thermodynamic definition. In (\cite{jpb3}), (\cite{jpb4}) it has 
been shown that we have 
\begin{equation} 
\frac {1}{T} = \frac{k_{B} \tau}{\hbar} +\frac{k_{B}}{\hbar}[U + \hbar 
\frac{d}{d \tau} \ln Z] \frac{d \tau}{dU} 
\label{defto} 
\end{equation}
that we can rewrite as  
\begin{equation} 
\frac{\hbar}{k_{B} T} = 
\tau + [ U - \smallint dx [<u_{K}(x)>_{path}+ <u_{P}(x)>_{path}]\frac{d 
\tau}{dU} 
\label{to} 
\end{equation}
in which $<u_{K}(x)>_{path}$ is the regular part of the mean value of the 
kinetic energy calculated over the paths localized around the 
initial point $x$ and $<u_{P}(x)>_{path}$ is a similar quantity but 
associated to the potential energy coming from the external potential. From 
(\ref{to}) we may deduce a natural equilibrium condition
\begin{equation} 
U = \smallint dx [<u_{K}(x)>_{path}+ <u_{P}(x)>_{path}]
\label{equi}
\end{equation}
It means that the mean value of the energy calculated on the paths 
corresponds to the internal energy needed to create the system. This 
equilibrium condition is very similar to the one used in (\cite{rove1}), it 
does not imply a time but as the consequence it leads to a time scale since 
from (\ref{to}) we get $\tau = \beta \hbar$ that is related to the 
temperature but not to the particular properties of the system as the mass 
of the particle or the external potential. When we introduce this 
value of $\tau$ into the expression of the path entropy we recover the 
thermodynamic result. Thus there is two equivalent definitions of the 
entropy; starting from the usual form of the partition function we recover 
the Boltzmann definition that we can relate to a number of states but from 
(\ref{gam}) we may also define the entropy as the disorder in space time.
This second definition is closely connected with the Gibbons Hawking 
(\cite{gibbons}) approach of the black holes thermodynamics in which 
the partition function is related to the euclidean action of the 
gravitational field and this action is associated with a time interval 
$\beta\hbar$. From our definition of entropy, it is also 
possible to establish a relation between action and entropy (\cite{jpb4}), a 
relation that is verified in the case of black holes. Thus the definition of 
entropy as the disorder in space time is equivalent to the Boltzmann one but 
may have a validity in more extended situations. \\ Clearly $\tau$ appears 
as the unit of time characterizing the thermodynamic equilibrium. From 
standard textbooks in statistical mechanics (\cite{landms}) it is well known 
that there is no entropy on a short period of time. Here there is no 
thermodynamics for time intervals smaller than $\tau$ since for $0 \le{t} 
\le{\tau}$ the quantum fluctuations are larger than the thermal ones having 
$\frac{1}{\beta}$ as order of magnitude. Thus to describe the thermodynamic 
evolution any time interval must be considered as a multiple of $\tau$ that 
appears as the natural unit of time.\\ The dynamics associated with the 
partition function allows us to derive a $\bf{H}$-$theorem$ as illustrated below.

\section{Derivation of a $\bf{H}$-$theorem$}
In Section 3 we have seen that the evolution in space time is characterized 
by the transition function $\phi(t,x)$ which is the solution of (\ref{dif}) 
verifying the initial condition $\phi_{0}(x)$. We have shown that the 
calculation of thermodynamic equilibrium properties requires to explore some 
loops during a time $\tau$. To this time interval we may associate a length 
$\Lambda =(\textbf{D} \tau)^{\frac{1}{2}}$ that is the thermal de Broglie 
wavelength multiplied by a numerical factor; $\textbf{D}=\frac{\hbar}{2m}$. 
Other lengths in the problem are the range of the external potential and, 
$L$, the one dimensional extension of the system. However $\phi(t,x)$ allows 
to investigate physical properties more general than just the equilibrium 
ones.\\ 
Let consider the conditions in which Boltzmann derived the so called 
$\bf{H}$-$theorem$. A system of particles is prepared in a given state by an 
external constraint. At $t = t_{0}$ the particles are distributed in space 
according to a distribution function $\phi_{0}(x)$ that is positive and 
normalized. At a given time $t=t_{0}$ the constraint is removed $i.e.$ the 
external potential is switched off, the system becomes free and relaxes 
towards an equilibrium state. The second law of thermodynamics asserts that 
during the relaxation it exists a given function that increase monotically 
with $(t-t_{0})$ and tends to the thermal entropy when $(t-t_{0}) \to 
\infty$.\\  Hence for $t \ge t_{0}$, $\phi(t,x)$ is the solution of 
\begin{equation} -{\partial  \phi(t,x)}/{ \partial t} + 
{\frac{\hbar}{2m} \Delta_{x} \phi(t,x)} = 0 
\label{diffree} 
\end{equation} 
In a one dimensional system of extension, $L$,  
the solution of (\ref{diffree}) is, taking $t_{0} = 0$,  
\begin{equation}
\phi(t,x) = \frac{1}{L} + \Sigma A_{n} \exp{-(\frac{\pi n\Lambda}{L})}^{2} 
\frac{t}{\tau}) \cos{(\frac{\pi n\Lambda}{L})}\frac{x}{\Lambda}
\label{fidis}
\end{equation}
on which the summation runs from $n=1$ to $n=\infty$ and the $A_{n}$ are the 
Fourier coefficients in the expansion of $\phi_{0}(x)$. $\phi(t,x)$ is 
normalized and it reduces to $\frac{1}{L}$ in the limit $t \to \infty$, it 
gives the probability of being in $x$ at time $t$. Now we may introduce a 
new definition of the entropy taking into account the evolution of 
$\phi(t,x)$. The equilibrium entropy defined in (\ref{gam}) can be 
written as $S_{eq} = k_{B} \ln (L \Gamma (\tau))$ with 
\begin{equation} 
\Gamma (\tau) = k_{B} \smallint Dx(t) \exp{ - \frac{1}{\hbar} [A_{o}[x(t);0, 
\tau] - \tau U] }. \label{gameq}
\end{equation}  
we may rewrite (\ref{gameq}) as 
\begin{equation}
S_{eq}= k_{B} \smallint dx \frac{1}{L}\ln (L \Gamma (\tau))
\label{gameqnew}
\end{equation}
That we can generalize by replacing $\frac{1}{L}$ by $\phi(t,x)$, we get 
\begin{equation}
S(t)= k_{B} \smallint dx \phi(t,x) \ln {\frac{\Gamma(\tau,t)}{\phi(t,x)}}
\label{snew}
\end{equation} 
In (\ref{snew}) the entropy $S(t)$ has two origins. One, related to 
$\phi(t,x)$, is associated with the evolution of the overall system and 
therefore with open paths. Smaller is $\phi(t,x)$ larger is the matter 
distribution and larger is $\ln{\frac{1}{\phi(t,x)}}$ $i.e.$ the 
contribution to $S(t)$. The second contribution is determined by 
$\Gamma(\tau,t)$ that is a generalization of $\Gamma(\tau)$ when the 
probability of space occupation is given by $\phi(t,x)$. This contribution 
is local and related to close paths. Thus, $S(t)$ defined by (\ref{snew}) 
contains the interplay between a global and a local relaxation. Let assume 
that we are in the thermodynamic limit $\frac{\Lambda}{L} \to 0$ and 
focusing on $t >> \tau$. We may expand $\phi(t+\tau,x)$ according to 
\begin{equation} \phi(t+\tau,x)= \phi(t,x) +  \frac{\partial 
\phi(t,x)}{\partial \frac{t}{\tau}} + ... 
 =  \phi(t,x) - \frac{\partial^{2}\phi(t,x)}{\partial 
 \frac{x^{2}}{\Lambda^{2}}} + ...
\label{expand}
\end{equation}
It is easy to calculate the partial derivatives using the explicit 
expression of $\phi(t,x)$ given in (\ref{fidis}). For $n$ 
finite $\frac{\partial^{2}\phi(t,x)}{\partial 
 \frac{x^{2}}{\Lambda^{2}}}$ goes to zero due to the thermodynamic limit, 
while if $(\frac{\pi n \Lambda}{L})$ becomes very large the partial 
derivatives are cancelled due to the condition $\frac{t}{\tau} >> 1$. Thus, 
there are realistic conditions for which the two contributions to $S(t)$ are 
uncoupled. Let consider now the function $\bf{H}(t)$ defined by $\bf{H}(t)= 
(\frac{1}{k_{B}}) (S(t) - S_{eq})$, it is given by \begin{equation}
\bf{H}(t)= - \smallint dx \phi(t,x) \ln {\phi(t,x)} - \ln{L}
\label{H}
\end{equation}
It is easy to verify (\cite{jpb4}) that $\bf{H}(t)$ is a monotonic 
function of time. Hence, in the thermodynamic limit, $\frac{\Lambda}{L} \to 
0$, and provided we inspect the system for times much larger than $\tau$ we 
have been able to define a monotonic increasing function of time $S(t)$ for 
which the limit $t \to \infty$ corresponds to the equilibrium entropy. The 
condition $t >> \tau$ leads to analyze the system for time interval $t$ on 
which the thermal fluctuations are much larger than the quantum ones, it 
corresponds to thermodynamic regime; at room temperature $\tau$ is 
approximately $20$ femtoseconds. 
 \\ Starting from the the transition function $\phi(t,x)$ it was easy to 
derive a $H-theorem$ because $\phi(t,x)$ is the solution of a time 
irreversible equation (\ref{dif}). Now to be really convincing we have to 
show that our approach can be extended in a natural way in order to describe 
reversible behaviors. This has been done in (\cite{jpb4}) where it has been 
shown that we may introduce a second transition function to describe the 
reverse motion. The two transition functions can be mixed into a complex 
valued function that verifies a Schroedinger like equation (\cite{jpb4}), 
more details will be given in (\cite{jpb5}) .\\

\section{Conclusions}
From the Gibbs expression of the density matrix several approaches have been
developed. Starting from an algebraic approach of quantum mechanics the 
Tomita-Takesaki theorem shows the existence of a one parameter isomorphism 
group. By comparing this result with the $KMS$ condition it seems natural to 
identify the parameter to a rescaled time. Then the partition function 
contains all the dynamics of the system for time larger than $\beta \hbar$. 
Starting from the path integral version of the partition function we may 
also discover a motion in space time but this one is restricted to a time 
interval smaller than $\beta \hbar$. If the dynamics issued of the 
Tomita-Takesaki theorem is generated by the hamiltonian in the path integral 
approach the dynamics is associated with a time irreversible equation. We 
have shown that it exists an equilibrium condition not based on a time 
condition but leading to introduce a time scale $\beta \hbar$ independent of 
the system properties like the mass or the external potential. The 
introduction of the equilibrium condition is similar to what has been done 
in (\cite{rove1}). We have shown that a $\bf{H}$-$theorem$ can be derived in the 
thermodynamic limit provided we focus on times larger than $\beta \hbar$.
We may conclude that the Feynman's conjecture leads to new aspects in 
statistical physics as expected by Feynman.

%%%%%%%%%%%%%%%%%%%%%%%%%%%%%%%%%%%%%%%%%%%%%%%%%%

\end{document}